\documentclass[twocolumn,secnumarabic,amssymb, nobibnotes, aps, prc,nofootinbib,fleqn, floatfix,showpacs,superscriptaddress]{revtex4-1}
\usepackage[utf8]{inputenc}
\usepackage{graphicx}
\usepackage{subfigure}
\usepackage{amsmath,amssymb}
\usepackage{bm}
\usepackage{multirow}

\begin{document}

\title{Global calculations on the microscopic energies and nuclear deformations: Isospin dependence of the spin-orbit coupling}%

\author{Zhe-Ying Wu}%
\affiliation{Department of Physics, Royal Institute of Technology (KTH), SE-10691 Stockholm, Sweden}

\author{Chong Qi}%
\email{chongq@kth.se}
\affiliation{Department of Physics, Royal Institute of Technology (KTH), SE-10691 Stockholm, Sweden}

\author{Ramon Wyss}%
\affiliation{Department of Physics, Royal Institute of Technology (KTH), SE-10691 Stockholm, Sweden}

\author{Hong-Liang Liu}%
\affiliation{Department of Applied Physics, School of Science, Xi'an Jiaotong University, Xi'an 710049, China}

\date{\today}%

\begin{abstract}
\begin{description}
\item[Background] The deviation between different model calculations that may occur when one goes toward regions where the masses are unknown is getting increased attention. This is related to the uncertainties of the different models which may have not been fully
understood.
\item[Purpose] 
To explore in detail the effect of the isospin dependence of the spin-orbital force in the Woods-Saxon potential on global binding energy and deformation calculations. 
\item[Method] The microscopic energies and nuclear deformations of about 1850 even-even nuclei are calculated systematically within the macroscopic-microscopic framework using three Woods-Saxon parameterizations, with different isospin dependences, which were constructed mainly for nuclear spectroscopy calculations. 
Calculations are performed in the deformation space $(\beta_2, \gamma, \beta_4)$.  Both the monopole and doubly stretched quadrupole interactions are considered for the pairing channel.
\item[Results] The ground state deformations obtained by the three calculations are quite similar to each other. Large differences are seen mainly in neutron-rich nuclei and in superheavy nuclei. Systematic calculations on the shape-coexisting second minima are also presented. As for the microscopic energies of the ground states, the results are also very close to each other. Only in a few cases the difference is larger than 2 MeV. The total binding energy is estimated by adding the macroscopic energy provided by the usual liquid drop model with its parameters fitted through the least square root and minimax criteria. Calculations are also compared with the results of other macroscopic-microscopic mass models.
\item[Conclusions] All the three calculations give similar values for the deformations, microscopic energies and binding energies of most nuclei. One may expect to have a better understanding of the isospin dependence of the spin-orbital force with more data on proton- and neutron-rich nuclei. 
\end{description}

\end{abstract}

\maketitle
%\tableofcontents

\section{INTRODUCTION}
Nuclear physics is an emergent phenomenon where regular and simple patterns can be created by the complicated interplay among its constitutes: protons and neutrons \cite{row10}. 
The understanding of its emergent behavior progresses by systematic experimental observations and the construction
of models to interpret them. New physics has been continuously revealed through the understanding of the global smooth behaviors of nuclear structure as well as its local fluctuations. 
 In particular, studies on the nuclear mass and other ground state properties reveal   strikingly systematic  behaviors including the nuclear liquid-like property, shell structure, pairing correlation and superfluidity as well as the nuclear deformation. It is not surprising that the so called macroscopic-microscopic (mac-mic) model \cite{Stru67,scw87,Naz85,Naza90,Moller1995185}, which is in line with above picture, has been extremely successful in describing the nuclear ground state properties and spectroscopy. 
 A reliable and precise theoretical prediction is also of particular importance for the estimation of the 
masses of experimentally unknown nuclei far from stability, particular those along the astrophysical
r-process path.  The mac-mic model and its derivations are still extensively applied nowadays \cite{Moller1995185,MollerPRL,wangning14,wangning13,
wangning11,wangning101,Zhang201438,MendozaTemis200828,MendozaTemis200884, PhysRevC.81.044321,PhysRevC.86.044316,Barbero201281,PhysRevC.86.021303,PhysRevC.89.024311, PhysRevC.67.044316,ian14,PhysRevC.90.064306,Royer201024,Royer20131}. 
With phenomenological corrections included, microscopic nuclear density functional calculations  can also give a comparable description of the binding energy \cite{Goriely201568}. Beyond mean-field corrections are also considered in recent global calculations \cite{arXiv:1407.7699,arXiv1502.06908}.
Besides these global calculations, local mass formulas and the nuclear shell model can give even more precise description for a limited number of nuclei (see, e.g., Refs. \cite{PhysRevC.90.064304, PhysRevC.90.054320, PhysRevC.86.044323} and references therein).

One perspective that is getting increased attention is the deviation between different effective theories that may occur when one goes toward regions where the masses are unknown. This is related to the uncertainties of the different models which may have not been thoroughly understood \cite{0954-3899-41-7-074001,0954-3899-42-4-045104}.
For example, a number of parameters of the Skyrme force still can not be fully determined by fitting to available experimental data and show large uncertainties \cite{PhysRevC.89.054314,PhysRevC.71.054311, arXive.1501.03572}. One encounters the same problem when determining the isospin dependence of the spin-orbit (SO) force of the phenomenological Woods-Saxon (WS) potential, which can have significant effects on the evolution of the shell structure in light neutron-rich nuclei  \cite{Xu2013247}. Such uncertainties can be related to the description of the single-particle structure in both self-consistent mean field and phenomenological approaches \cite{arXiv:1501.04148,Xu2013247}. The influence of the SO coupling of the Skyrme force on global binding energy calculations was recently studied in Ref. \citep{Goriely201568}. 

The motivation of this work is to calculate systematically the microscopic energies and nuclear deformations within the mac-mic framework with different WS parameterizations which were constructed primarily for nuclear spectroscopy calculations. In particular, we are interested to explore in detail the performance of the WS parameterization of Ref. \cite{Xu2013247} in heavy nuclei and to see whether the isospin dependence of the SO force has any global influence on binding energy and deformation. Calculations will also be compared with the results of other mac-mic mass models.

The paper is organized as follows: In Sec. II, we briefly introduce the mac-mic approach and the empirical WS single-particle potential. It is followed by the detailed comparisons on the calculated nuclear deformations and microscopic energies in Sec. III. 
The total binding energies are studied in Sec IV where the parameters of the liquid drop model are fitted to experimental binding energies. A short summary is given in Sec. V.

\section{The theoretical framework: Microscopic energy and the nuclear deformation}

Within the mac-mic framework \cite{Stru67}, the total energy of a nucleus can be written as the sum of a macroscopic and a microscopic terms.  The macroscopic term describes the bulk properties of the nucleus. It is usually approximated by the standard liquid drop model or its revised versions.  The microscopic term, which may show large fluctuations with changing deformation and particle number, has its origins in the quantum shell effects. It describes the single particle properties of the nucleons near the Fermi surface. It usually consists of  the shell and pairing correction terms, which are evaluated in an average potential well. Thus, for a given nucleus with Z protons and N neutrons at the full set of deformation parameters $\beta$, the total energy can be written as \cite{Naza90}:
\begin{eqnarray} 
E(N,Z,\beta)&=&E_{mac} (N,Z) + E_{def} (N,Z,\beta)\nonumber \\ 
&+&E_{shell} (N,Z,\beta)+ E_{pair}(N,Z,\beta)
\end{eqnarray} 
The deformation correction energy is written as \cite{Ld74,scw87}
  \begin{eqnarray} 
E_{def} =[B_s(\beta)-1]E_s^{(0)}+[B_c(\beta)-1]E_c^{(0)}
\end{eqnarray} 
where $B_s$ and $B_c$ are functions of the shape of the nucleus only and are equal to 1 when the nucleus is spherical. $E_s^{(0)}$ and $E_c^{(0)}$ refer to the spherical surface energy and the spherical Coulomb energy, respectively.

In this paper, the single-particle level is derived from a non-axial deformed WS potential \cite{ws54,Dud79,Dud81, Naz85,scw87,Wyss91} of the form:
\begin{eqnarray} 
V(\bm{r},\beta)&=&\frac{V}{1+exp[(\bm{r}-\bm{R})/a]}\nonumber\\
&+&\nabla\frac{V_{so}}{1+exp[(\bm{r}-\bm{R})/a_{SO}]}(\bm{\sigma} \times \bm{p})\nonumber\\
&+&\frac{1}{2}\left(1+\tau_3\right)V_{Coul}
\end{eqnarray} 
which correspond to the central potential, the SO potential and the Coulomb potential, respectively.  The surface is defined as
\begin{eqnarray} 
\bm{R}=C(\beta)R_0\left(1+\sum_{\lambda,\mu}{\alpha_{\lambda,\mu}Y_{\lambda,\mu}}\right)
\end{eqnarray} 
where $C(\beta)$ is the volume conservation factor.
The Lund convention has been used to transform the coefficients $\alpha_{\lambda,\mu}$ to nuclear deformations in terms of $\beta$ and $\gamma$. 
In the present work, a non-axially symmetric shape with coefficients up to $\beta_{4\kappa}$  is considered but only even multiples are included. That is, the potential energy surface calculation is performed in the deformation space ($\beta_2$,$\gamma$,$\beta_4$) \cite{Naz85}.
The ground state deformation values are taken as those that correspond to the minimum in the total energy. The influence of the octupole and higher-order deformations is considered later in a systematic calculation within a axially symmetric deformation space.

A variety of parameterizations of the WS potential exists (see, e.g., Refs. \cite{Bohr69,Blo60,Isa02,Dud82}, Table II in Ref. \cite{Sch07} and Table I in Ref. \cite{dud98}). In the ``standard" one \cite{Bohr69,Blo60}, the strengths of the central and SO potentials are given as
\begin{equation}
V=-V_0(1+\frac{4\kappa}{A}{\bf t}\cdot{\bf T}_{d}),
\end{equation}
and
\begin{equation}
V_{SO}=-\lambda V_0(1+\frac{4\kappa_{SO}}{A}{\bf t}\cdot{\bf T}_{d}),
\end{equation}
respectively, where we have replaced the original $N-Z$ term with $4{\bf t}\cdot{\bf T}_{d}$ to get a consistent description of both protons and neutron orbitals.
${\bf t}$ and ${\bf T}_{d}$ denote the isospin quantum numbers of the last nucleon and of the daughter nucleus, respectively. The total isospin of the system is ${\bf T}={\bf t} + {\bf T}_{A-1}$.
It is $4{\bf t}\cdot{\bf T}_{A-1}=-3$ for the $T=0$ ground state of a $N=Z$ nucleus and 
\begin{eqnarray}
\nonumber 4{\bf t}\cdot{\bf T}_{A-1}&=&N-Z-1 {\rm ~for~neutron~orbits}\\
&=&-(N-Z+3)  {\rm ~for~proton~orbits~~~~}
\end{eqnarray}
in $N> Z$ nuclei with $T=(N-Z)/2$ \cite{Sch07}. In Ref. \cite{Bohr69}, the isospin-dependent terms are parameterized as 
\begin{equation}
\kappa=\kappa_{SO}=-\frac{33}{51},
\end{equation}
where the SO potential
depth is assumed to have the same isospin dependence as that of the central
potential. This assumption is rather commonly used \cite{Dud82}. The typical strength of $\kappa$ is in the range $-0.6\sim-0.9$. $\kappa_{SO}$ is assumed to be zero in Ref. \cite{Sch07}. 
In Ref. \citep{Xu2013247}, an unconventional assumption is taken as
\begin{equation}
\kappa_{SO}=-\kappa
\end{equation}
in order to explain the shell evolution in light neutron-rich nuclei. We have shown that above choice of $\kappa_{SO}$ is essential for the description of the disappearence of the $N=8$, 20 shell closures and the emergence of the $N=6$, 14, 16, 32 as well as 34 new subshells in neutron-rich nuclei. A number of selected spherical as well as deformed nuclei are evaluated in Ref. \cite{Xu2013247} but its global performance has not yet been explored.
In particular, our calculations reproduce well the intruder configurations and ground state deformations of neutron-rich nuclei around $N=20$. A similar island of inversion is expected in neutron-rich nuclei around $N=40$.

The shell correction energy has been calculated by the traditional Strutinsky methods \cite{Stru75}. A correction polynomial of the order $p=6$ and the smoothing parameter $\gamma=1.2\hbar\omega_0$($\hbar\omega_0=41 A^{-1/3}$) has been used in this paper.  The pairing energy is calculated considering both the monopole and doubly stretched quadrupole interaction:
\begin{eqnarray}
\overline{v}_{\alpha\beta\gamma\delta}^{(\lambda\mu)}=-G_{\lambda\mu}g_{\alpha\overline{\beta}}^{(\lambda\mu)}g_{\gamma\overline{\delta}}^{*(\lambda\mu)}
\end{eqnarray} 
where
\begin{eqnarray}
g_{\alpha\overline{\beta}}^{(\lambda\mu)}=\left\{
\begin{array}{rcl}
&&\delta_{\alpha\overline{\beta}}~~~~~~~~~~~~~~~{\lambda=0,   \mu=0, } \\
&&\langle\alpha|\widetilde{Q}_{\mu}|\overline{\beta}\rangle ~~~~~~~~ {\lambda=2,   \mu=0, 1, 2.}
\end{array}
\right.
\end{eqnarray} 
Here, $\alpha$($\overline{\alpha}$) denotes the states of signature $r=-i$ ($r=-i$). Using the double-stretched operator ($Q_\mu^{\prime\prime}=r^{\prime\prime 2}Y_{2\mu}^{\prime\prime}$), the generators of quadrupole pairing interaction in Eqs. (11) have the form:    
\begin{eqnarray}
&&\widetilde{Q}_0=Q_{20}^{\prime\prime},\nonumber\\
&&\widetilde{Q}_\mu=\frac{1}{\sqrt{2}}(Q_{2\mu}^{\prime\prime}+Q_{2-\mu}^{\prime\prime}),~~~~~~~~~~\mu=1,2.
\end{eqnarray}
The monopole paring strength, $G_{00}$, is determined by the average gap method\cite{Moller92}, while the quadrupole pairing strengths, $G_{2\mu}$, are obtained by restoring the Galilean invariance of the system under $\lambda$-pole collective shape oscillations \cite{pair90}. 
In order to avoid the spurious pairing phase transition and particle number fluctuation encountered in the BCS calculation,  the pairing is treated by the Lipkin-Nogami approach \cite{LN73}  in which the particle number is conserved approximately. For further details, see Refs. \cite{satula94,satuc94,satula95,Xu2000119}.

\section{Global calculations}
We have performed systematic calculations for all even-even nuclei with $Z,N\geq8$ within the proton and neutron driplines as defined in Fig. 12 in Ref. \cite{0954-3899-42-4-045104}.  In total 1871 nuclei are calculated and the ground state is defined by taking the point with the minimum energy within the deformation space ($\beta_2$,$\gamma$,$\beta_4$).
We focus in this paper on the detailed comparison for known nuclei and thus do not explore the positions of the driplines as predicted by the present model, which may depend on the choice of the macroscopic energy model and the uncertainties induced by the pairing interaction \cite{arXiv:1501.04148,Phyc}.
The calculated potential energy surfaces for all above nuclei as well as the deformations and microscopic energies corresponding to the global minimum will be available on our web page \cite{web}.

\begin{table*}
\caption{\label{tab:tableWS} Parameters of the three Woods-Saxon potentials employed in our calculations. For WS1 and WS3 one has $R_0=r_0A^{1/3}$ whereas for WS2 it is $R_0=r_0(1+0.116\tau_3(N-Z)/A)A^{1/3}+0.235$fm and $R_{SO}=R_0 $.}
\begin{tabular}{cccccccc}
\hline
\hline
&V$_0$(MeV)& r$_0$(fm)&r$_{SO}$(fm)&a,a$_{SO}$(fm)&$\lambda$&$\kappa$&$\kappa_{SO}$\\
\hline
WS1& 49.6 &1.347(n)/1.275(p) & 1.31(n)/1.32(p)&0.7&35(n)/36(p)&0.86&$\kappa$\\
WS2& 53.75&1.19&1.19&0.637&29.49& 0.791&0.162\\
WS3&50.92&1.285&1.146&0.691&24.08&0.644&$-\kappa$\\
\hline
\hline
\end{tabular}
\end{table*}

Three WS parameterizations \cite{Dud82,wyss,Xu2013247} are employed in our calculations, which were constructed for different purposes. The so-called Universal parameter set has been extensively used in both nuclear structure (e.g., Refs. \cite{PhysRevC.89.044304,PhysRevC.87.044319,PhysRevLett.92.252501,Liu07})and radioactive decay \cite{Delion95,Qi12,Qi10,Kar06} studies. The parameter set from Ref. \cite{wyss} (see Table \ref{tab:tableWS} below and Sec. IIIC in Ref. \cite{PhysRevC.81.044321}) has been shown to be very successfully in both nuclear spectroscopy \cite{PhysRevC.90.064309,PhysRevC.87.057302, PhysRevC.86.044302, PhysRevC.85.027301} and binding energy \cite{PhysRevC.81.044321} calculations. In particular, it gives a good description of the nuclear momentum of inertia and high spin states. 
We would also like to emphasize that above parameter set assumes a rather sophiscated radius of the form $R_0=r_0(1+0.116\tau_3(N-Z)/A)A^{1/3}+0.235$ fm. 
The parameter set from Ref. \cite{Xu2013247} contains much less terms and is primarily fitted to doubly-magic nuclei with the restriction that the radius is smaller than $1.3A^{1/3}$. This potential has not been much tested. We are also interested to see whether it has any major defects in global nuclear structure calculations. We label the three parameter sets as WS1, WS2, WS3 for simplicity. This should not be confused with the WS mass formula as proposed in Ref. \cite{wangning101}.  WS1 has a very strong isospin dependence in SO force with  $\kappa=\kappa_{SO}=0.86$. It is much weaker in WS2, where $\kappa_{SO}$ is a free parameter with the value of  0.16.
The parameters of the three WS potentials are listed in Table \ref{tab:tableWS}.

\subsection{Systematics on nuclear deformation}

\begin{figure}
\includegraphics[width=0.5\textwidth]{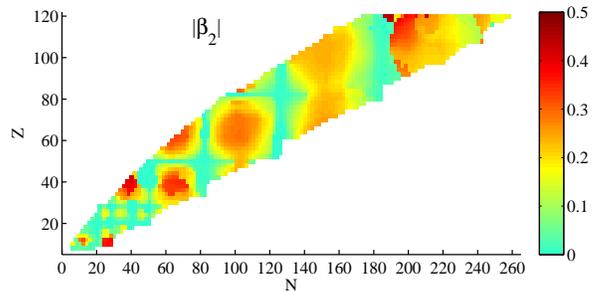}
\caption{\label{duniv}(Color online) Systematic calculations on the quadrupole deformation $\beta_2$ of even-even nuclei by using the Woods-Saxon potential with WS1 from Ref. \cite{Dud82}.}
\end{figure}

\begin{figure}
\includegraphics[width=0.5\textwidth]{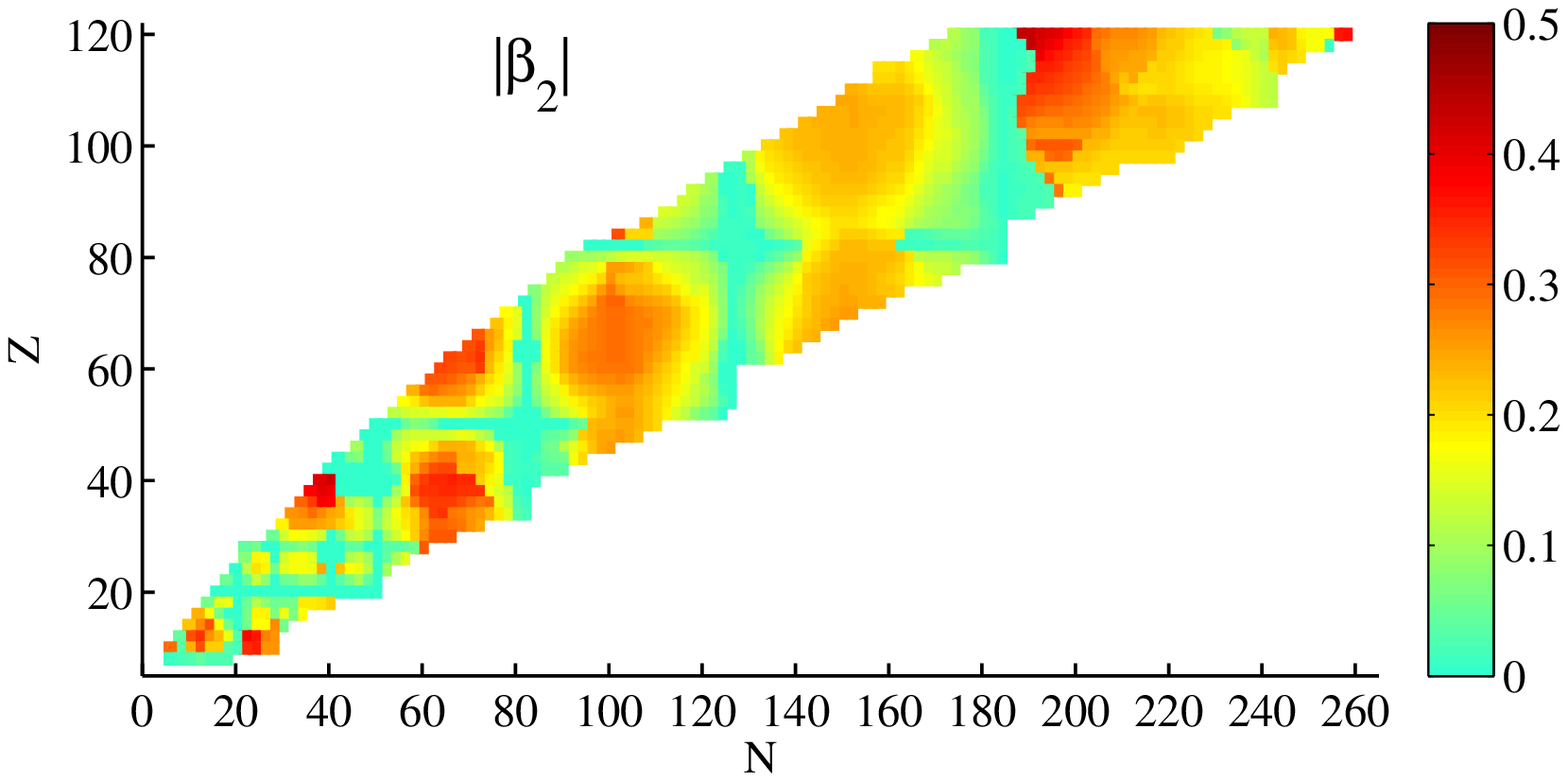}
\caption{\label{dcrank}(Color online)  Same as Fig. \ref{duniv} but for calculations with WS2 from Ref. \cite{wyss}.}
\end{figure}

\begin{figure}
\includegraphics[width=0.5\textwidth]{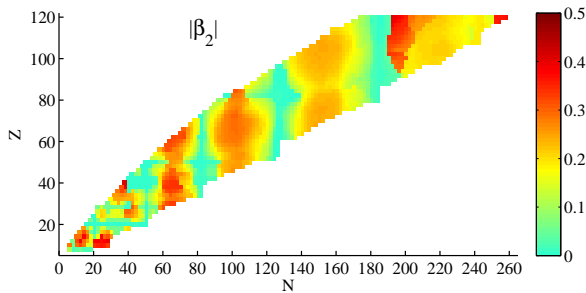}
\caption{\label{dxu} (Color online)  Same as Fig. \ref{duniv} but for calculations with WS3 from Ref. \cite{Xu2013247}.}
\end{figure}

The calculated quadrupole deformations $\beta_2$ with the different WS parameter sets are plotted in Figs. \ref{duniv}, \ref{dcrank} and \ref{dxu}. $\beta_2$ is assumed to be positive. A prolate (oblate) shape corresponds to $\gamma\sim0 (60)\deg$. The nucleus would be of maximal triaxial deformation if one has 
$\gamma\sim30 \deg$. The $\gamma$ deformation was not taken into account in some recent mass-formula calculations. Its importance is emphasized in Ref. \cite{PhysRevLett.97.162502}.

\begin{figure}
\includegraphics[width=0.3\textwidth]{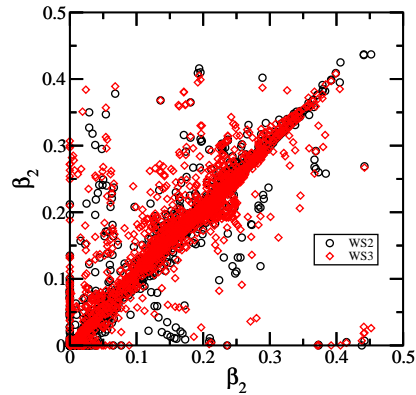}
\caption{\label{diffb} (Color online)  Comparison between the deformations calculated from the three different WS parameters. We plot the results from parameters WS2,WS3 as a function of those from WS1.}
\end{figure}

In most cases, as can be seen from Fig. \ref{diffb},  the deformations calculated by the three parameters are quite close to each other.
For calculations with parameters WS1 and WS2, there are only 72 cases where differences between the quadrupole deformation $\Delta\beta_2$ are larger than 0.1, among which one has 20 with $\Delta\beta_2\geq0.2$ and 7 with $\Delta\beta_2\geq0.3$. The latter correspond to nuclei $^{34}$Mg, $^{80}$Sr, $^{82}$Zr, $^{82}$Mo, $^{292}$Cn, $^{306}$118, $^{306}$120. For  the WS1 and WS3 parameters, there are 118 cases where differences between the quadrupole deformation $\Delta\beta_2$ are larger than 0.1, among which one has 50 (17) with $\Delta\beta_2\geq0.2 (0.3)$.  As discussed in Ref. \cite{Xu2013247}, many of them are neutron-rich nuclei with $N$ around 20 and 40 due to the vanishing of the shell closure. As can be seen in Fig. \ref{dxu}, for calculations with WS3, the ground states of a few Sn and Pb isotopes are calculated to be deformed, which are slighter lower in energy than the spherical minima.

We also compared our calculated deformations with those given by the mass calculations of Refs. \cite{Moller1995185,wangning101}. For the deformations given by  WS3 parameter  and those from Ref. \cite{Moller1995185}, there are as many as 66 (34) cases with $\Delta\beta_2\geq0.2 (0.3)$. However, in the latter cases with $\Delta\beta_2\geq0.3$, 27 are in the superheavy region around $N=192$.
In comparison to those given in Ref. \cite{wangning101}, there are 30 cases where one has $\Delta\beta_2\geq0.2$.
Besides the few differences around $N=192$, the other cases are mainly around $N, Z=14, 28$ and 40. In particular, $^{78}$Zr is predicted to be largely prolate deformed in our calculation with WS3 but oblate deformed in Ref. \cite{wangning101}. $^{64}$Ge is predicted to be of triaxial shape with $\gamma=41\deg$ in our calculation with WS3.

\begin{figure}
\includegraphics[width=0.5\textwidth]{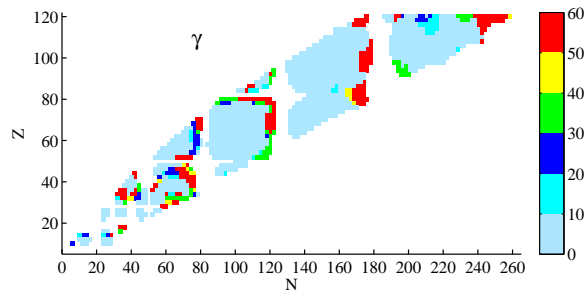}
\caption{\label{gammauniv} (Color online)  Systematic calculations on the gamma deformation $\gamma$ of even-even nuclei by using the Woods-Saxon potential with WS1. Only results corresponding to nuclei with $|\beta_2|>0.1$ are shown.}
\end{figure}

\begin{figure}
\includegraphics[width=0.5\textwidth]{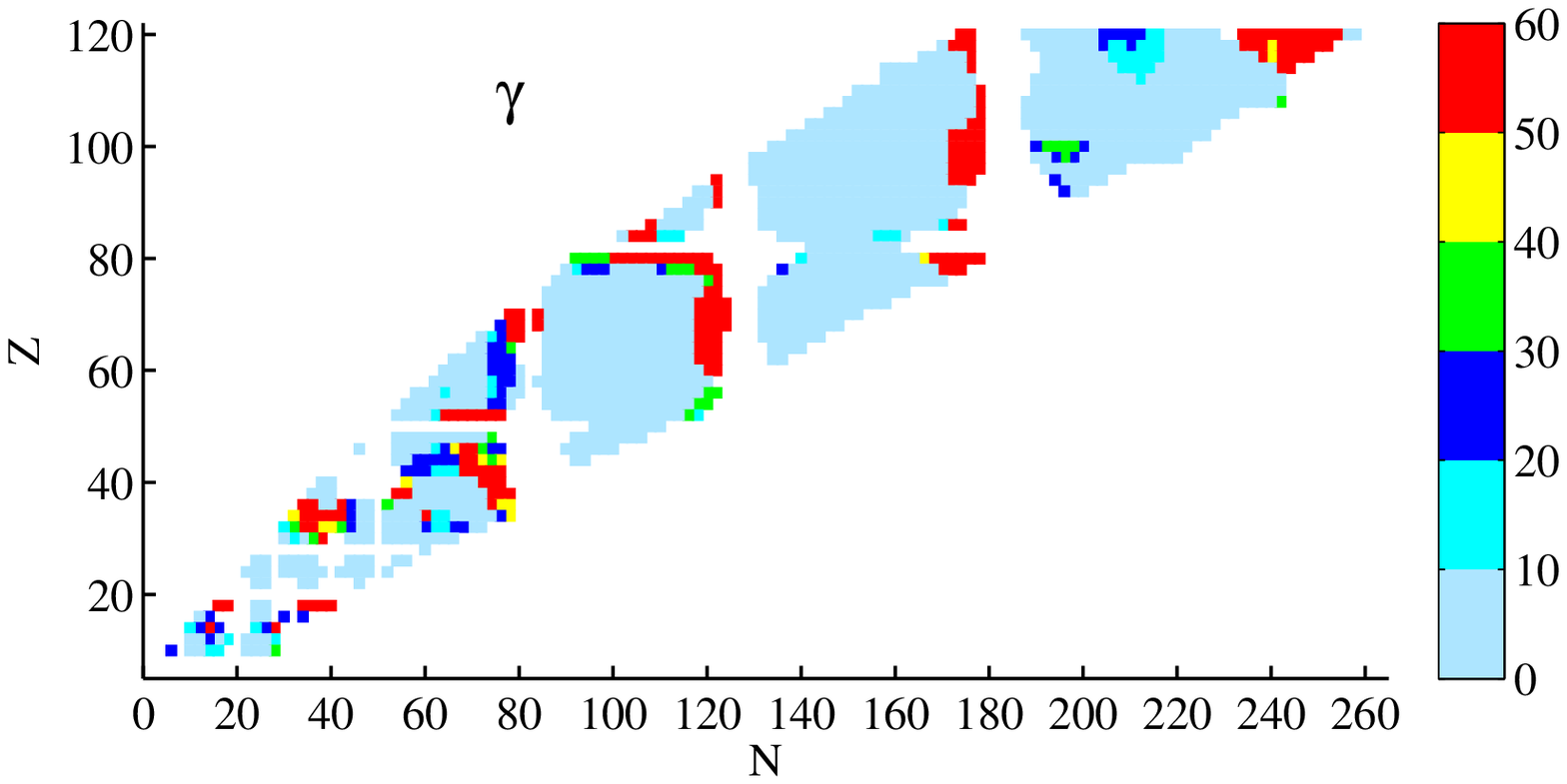}
\caption{\label{gammacrank}(Color online)  Same as Fig. \ref{gammauniv} but for calculations with WS2.}
\end{figure}

\begin{figure}
\includegraphics[width=0.5\textwidth]{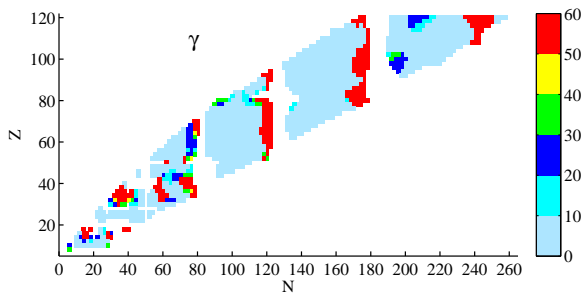}
\caption{\label{gammaxu} (Color online)  Same as Fig. \ref{gammauniv} but for calculations with WS3.}
\end{figure}

The gamma deformations as calculated with the different WS parameter sets are plotted in Figs. \ref{gammauniv}, \ref{gammacrank} and \ref{gammaxu}. 
Again, the three calculations give practically the same results for most nuclei. As can be seen from the figures, most nuclei are of prolate shape with $\gamma<10\deg$. The nuclei with triaxial and oblate deformations occur mostly in regions where transition from spherical to prolate shape is expected.  
In this context, it may be interesting to mention that there is a long history studying the origin of dominance of nuclei with prolate shape, which may be sensitive to the details of the WS parameterization \cite{PhysRevC.86.064323,Takahara2011429}. However, the predicted deformations seem pretty stable within the range of the variety of WS parameterizations employed in this work.

\subsection{Deformations of selected isotopes}

\begin{figure}
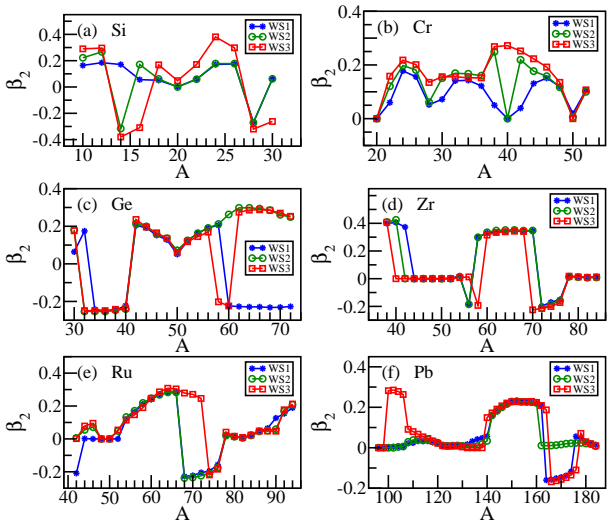

\includegraphics[width=0.22\textwidth]{isotope14.eps}
\includegraphics[width=0.22\textwidth]{isotope24.eps}
\includegraphics[width=0.22\textwidth]{isotope32.eps}
\includegraphics[width=0.22\textwidth]{isotope40.eps}
\includegraphics[width=0.22\textwidth]{isotope44.eps}
\includegraphics[width=0.22\textwidth]{isotope82.eps}
\caption{\label{i30} (Color online)  Deformations of Si, Cr,Ge, Zr, Ru and Pb isotopes as predicted by different WS calculations.}
\end{figure}

As can be inferred from Figs. \ref{duniv}-\ref{diffb}, the deformations predicted by the three calculations are quite similar in most cases. We are particularly interested in the differences between predictions by the three calculations.
In Fig. \ref{i30} we plotted the predicted deformations of Si, Cr, Ge, Zr, Ru and Pb isotopes.

For Si isotopes, the large difference occurs around $N=Z=14$. The ground state of $^{28}$Si is calculated to be of triaxial shape with $\beta_2=0.17$ and $\gamma=22^\circ$ by WS1 parameter set. It also gives another two energy minima: the second minimum is calculated to be of nearly spherical shape with $\beta_2=0.07$ and the third one is of oblate shape with $\beta_2=0.27$. The energy difference between these three minima is only about 100 keV. It can be inferred that the shape of $^{28}$Si predicted by the WS1 calculation is quite soft.  The ground state of this nucleus is calculated to be of oblate shape with $\beta_2=0.32$ and $0.38$ in calculations with WS2 and WS3, respectively. A coexisting second minimum with triaxial deformation is also predicted in these two calculations.  For calculation with parameters WS2(WS3), the energy difference between the two minima is about 300 keV (820 keV).

For Cr isotopes,  the largest difference occurs in neutron-rich nuclei around $N=40$. The nucleus
 $^{64}$Cr is predicted to be spherical in the calculation with the WS1 parameter, while 
it is predicted to be of well deformed prolate shape with  $\beta_2=0.27$ by the WS3 calculation.  For the calculation with the parameter set WS2, the coexistence of spherical and prolate shapes has been seen.

For Ge isotopes, large differences occur in nuclei with N$\ge$56. But for those nuclei, the ground state shapes calculated with the three WS parameter sets are rather soft in the $\gamma$ direction from $-30^\circ$ to $60^\circ$. The $\beta_2$ values are quite close to each other.

For Zr isotopes, the trends of the evolution of the nuclear shape with neutron number calculated with the three WS parameter sets are quite similar to each other. However, the locations of the transitions are slightly different. It can be seen that the transition from well deformed shape  to spherical shape around $N=40$ occurs earlier in the calculation with the WS3 parameter than in calculations by WS1 and WS2 parameters, while the evolution from spherical shape to oblate shape around $N=56$ occurs later in the WS3  calculations as compared to those by WS1 and WS2. 

For Ru isotopes, the coexistence of spherical  shape and oblate deformation  appears in nuclei around $N=42$.   The isotopes with $N=60-66$ are predicted to be triaxially deformed with $\beta_2\sim 0.25$ and $\gamma\sim24^\circ$. Then transition from triaxial to oblate deformation occurs at $N=68$ in calculations by WS1 and WS2 parameters while triaxial shape remains until $N=74$ in calculations with the WS3 parameter.

For Pb isotopes,  large differences occur in light lead nuclei around $N=102$ and in heavy nuclei around $N=170$. It is well known that the coexistence of spherical, prolate and oblate shapes occurs in light lead nuclei. The ground-state deformations of those nuclei are spherical according to calculations with the WS1 and WS2 parameter sets whereas the prolate minima are lower in energy than the spherical minima in calculations with the WS3 parameter. 
This may be due to the simplified treatment of the radius parameters in WS3.
For neutron-rich nuclei with $N\sim170$, the ground-state deformation is calculated to be spherical by WS2 parameter while oblate deformation is obtained by the WS1 and WS3 calculations.

\begin{figure}
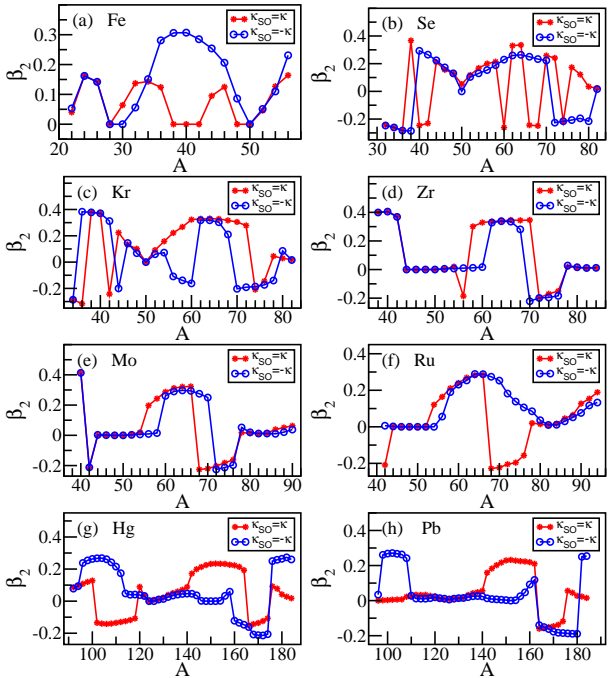

\includegraphics[width=0.220\textwidth]{isotope26univok.eps}
\includegraphics[width=0.220\textwidth]{isotope34univok.eps}
\includegraphics[width=0.220\textwidth]{isotope36univok.eps}
\includegraphics[width=0.220\textwidth]{isotope40univok.eps}
\includegraphics[width=0.220\textwidth]{isotope42univok.eps}
\includegraphics[width=0.220\textwidth]{isotope44univok.eps}
\includegraphics[width=0.220\textwidth]{isotope80univok.eps}
\includegraphics[width=0.220\textwidth]{isotope82univok.eps}
\caption{\label{uvok} (Color online)  Comparion of ground state deformations for selected isotopes with and without the flip in $\kappa_{SO}$ for calculations with WS1.}
\end{figure}

\begin{figure}
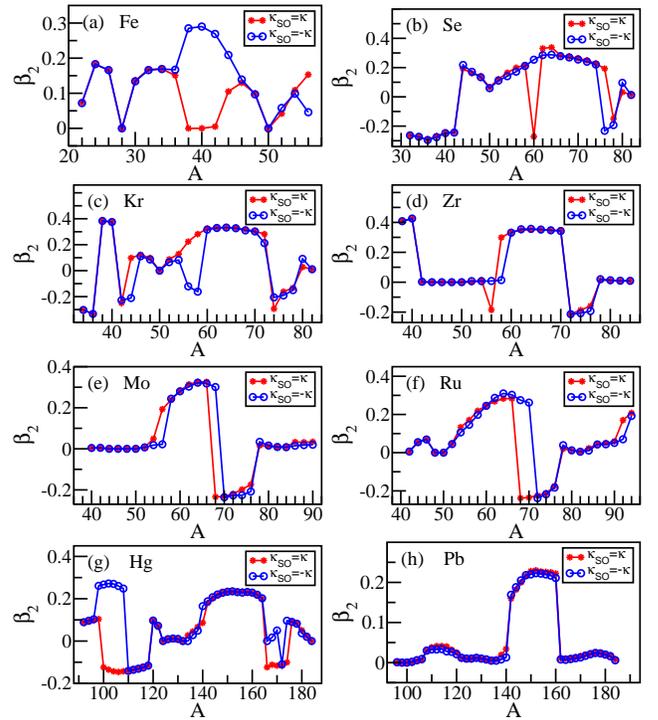

\includegraphics[width=0.230\textwidth]{isotope26crankok.eps}
\includegraphics[width=0.230\textwidth]{isotope34crankok.eps}
\includegraphics[width=0.230\textwidth]{isotope36crankok.eps}
\includegraphics[width=0.230\textwidth]{isotope40crankok.eps}
\includegraphics[width=0.230\textwidth]{isotope42crankok.eps}
\includegraphics[width=0.230\textwidth]{isotope44crankok.eps}
\includegraphics[width=0.230\textwidth]{isotope80crankok.eps}
\includegraphics[width=0.230\textwidth]{isotope82crankok.eps}
\caption{\label{crankok}(Color online)  Same as Fig. \ref{uvok} but for calculations with WS2.}
\end{figure}

The isospin-dependence of the SO force can indeed lead to large fluctuations in ground state deformations for certain nuclei. Its effect on light neutron-rich nuclei is discussed in Ref. \cite{Xu2013247}. To explore this point further, we have done schematic calculations with WS1 and WS2 by changing the sign of the isospin-dependence of the SO force $\kappa_{SO}$. In Figs. \ref{uvok} and \ref{crankok}, we plotted a few selected isotopic chains where the flip is expected to have large effects on the shape prediction. It is not surprising to see that the effect is larger in calculations with the WS1 parameter which has a strong isospin-dependent SO force than that in WS2. However, in the latter case, large deviations can also be seen in neutron-rich nuclei including those around $N=40$ and 60.  
 
\subsection{The calculated microscopic energies}

\begin{figure}
\includegraphics[width=0.5\textwidth]{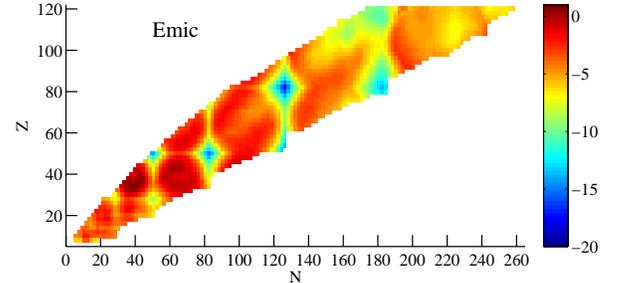}
\caption{\label{etotuniv} (Color online)  Systematic calculations on the total microscopic energies, $E_{def} (N,Z,\beta) + E_{shell} (N,Z,\beta)+ E_{pair}(N,Z,\beta)$, of even-even nuclei by using the Woods-Saxon potential with WS1.}
\end{figure}

\begin{figure}
\includegraphics[width=0.5\textwidth]{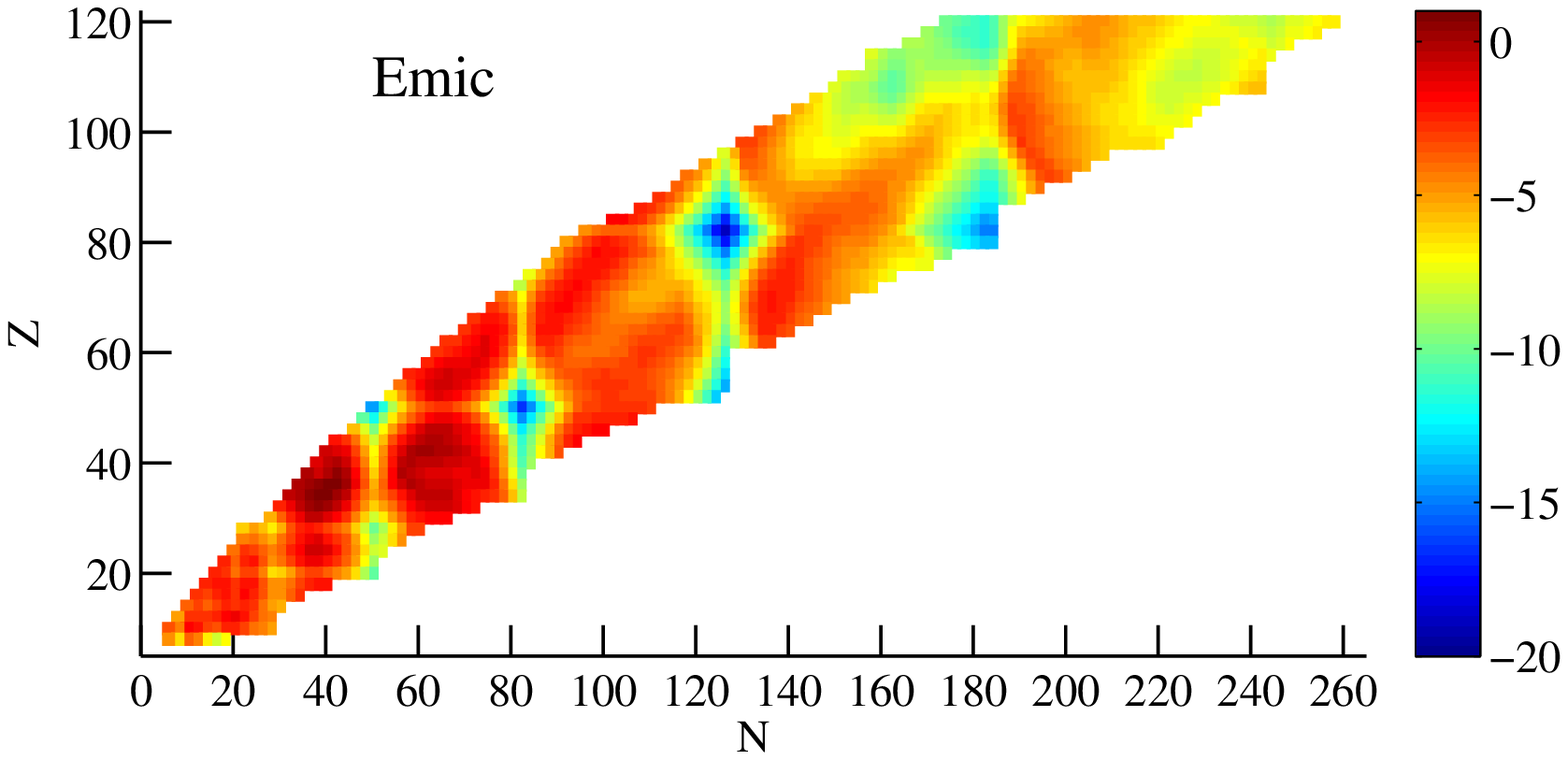}
\caption{\label{etotcrank}(Color online)  Same as Fig. \ref{etotuniv} but for calculations with WS2.}
\end{figure}

\begin{figure}
\includegraphics[width=0.5\textwidth]{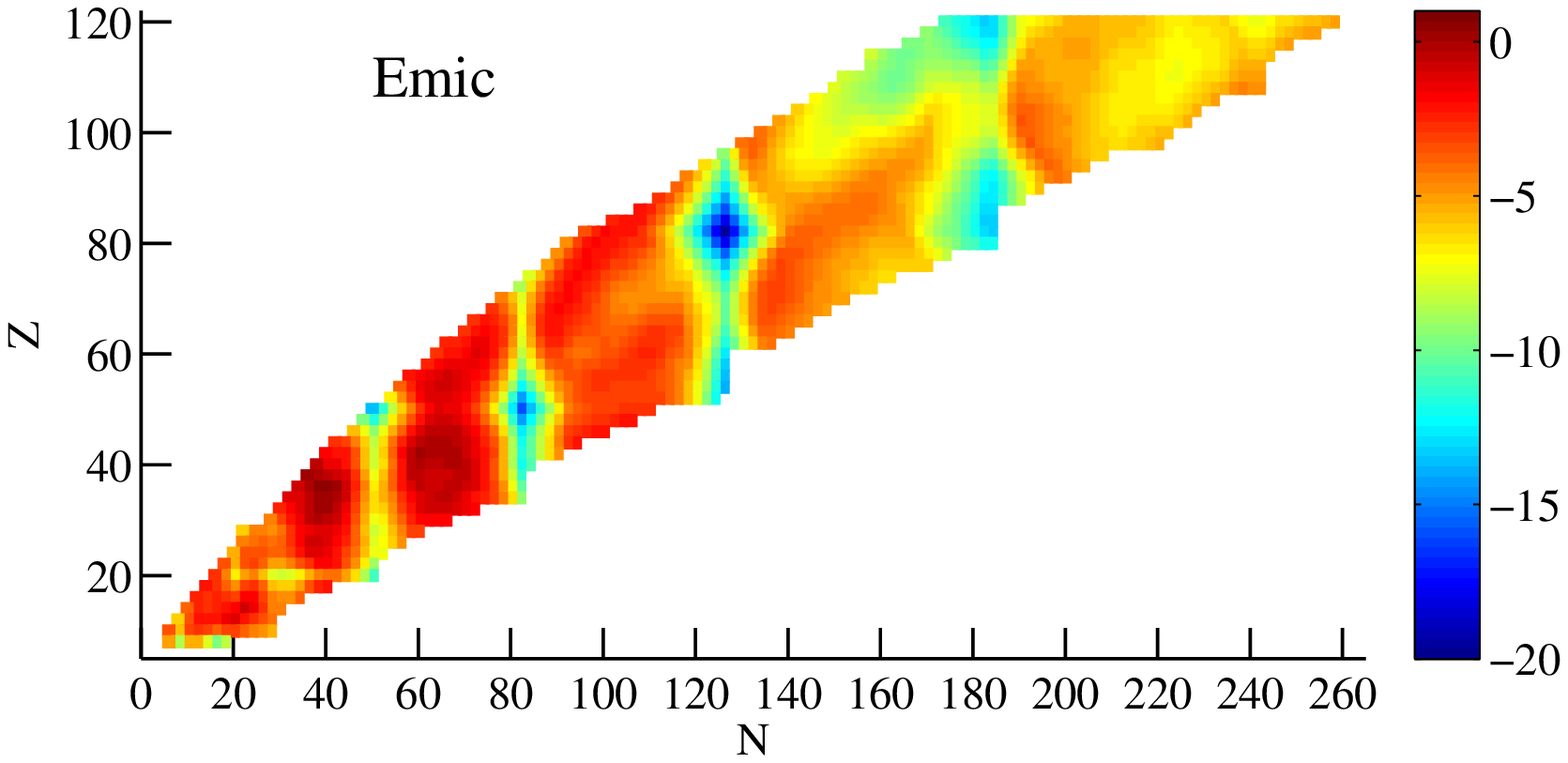}
\caption{\label{etotxu} (Color online)  Same as Fig. \ref{etotuniv} but for calculations with WS3.}
\end{figure}

The microscopic energy depends on the deformation as well as the single-particle structure of the nucleus to be studied. Even though it is not a direct observable, the microscopic energy can provide an interesting test to our single-particle potentials. In Figs. \ref{etotuniv}, \ref{etotcrank} and \ref{etotxu} we plotted the total microscopic energy, $E_{def} (N,Z,\beta) + E_{shell} (N,Z,\beta)+ E_{pair}(N,Z,\beta)$, for all three calculations as a function of $N$ and $Z$.

\begin{figure}
\includegraphics[width=0.3\textwidth]{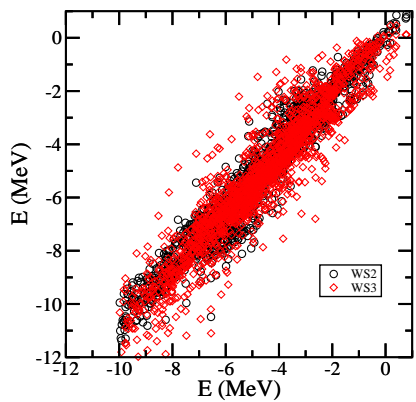}
\caption{\label{cc} (Color online) Comparison between calculated total microscopic energies from the three calculations. We plot the results from parameters WS2, WS3 as a function of those from WS1.}
\end{figure}

For most cases, the three calculations give similar description on the microscopic energy, as can be seen in Fig. \ref{cc}. 
For the two calculations with parameters WS1 and WS3, there are in total 124 cases that the differences are larger than 2 MeV.
For calculations with parameters WS3 and WS2, there are 23 cases  with differences larger than 2 MeV.
For calculations with parameters WS1 and WS2, there are 28 cases with differences larger than 2 MeV.

\begin{figure}
\includegraphics[width=0.5\textwidth]{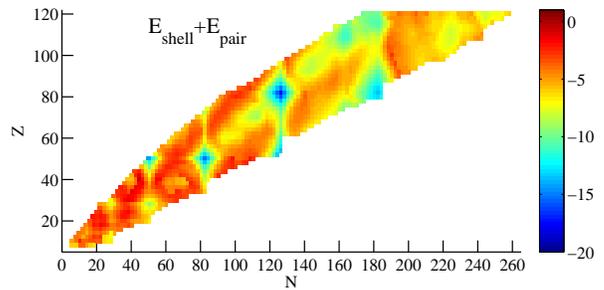}
\caption{\label{emicuniv} (Color online) Systematic calculations on the shell plus pairing correction energy of even-even nuclei by using the Woods-Saxon potential with WS1.}
\end{figure}

\begin{figure}
\includegraphics[width=0.5\textwidth]{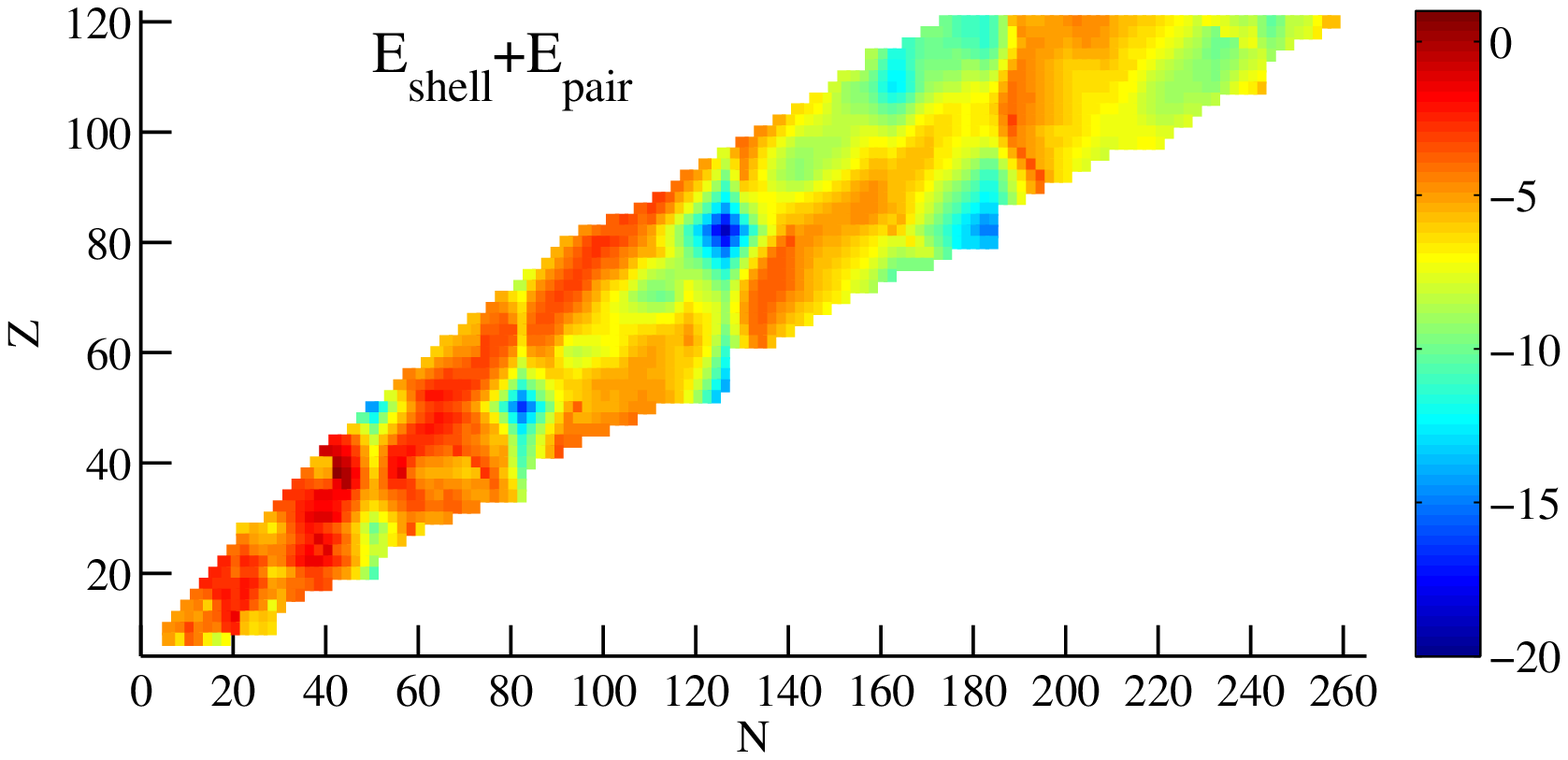}
\caption{\label{emiccrank} (Color online)  Same as Fig. \ref{emicuniv} but for calculations with WS2.}
\end{figure}

\begin{figure}
\includegraphics[width=0.5\textwidth]{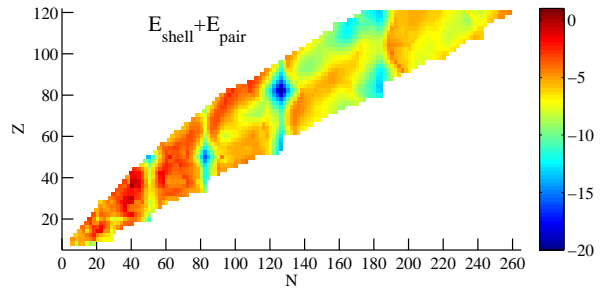}
\caption{\label{emicxu} (Color online)  Same as Fig. \ref{emicuniv} but for calculations with WS3.}
\end{figure}

In some studies, only the shell correction and pairing correction energies are considered. The sum of these two corrections are plotted in Figs. \ref{emicuniv}, \ref{emiccrank} and \ref{emicxu}. Comparing the two calculations with parameters WS3 and WS1, there are in total 163 cases that the differences are larger than 2 MeV,
among which there are 7 cases with differences larger than 4 MeV.
For calculations with parameters WS3 and WS2, there are 48 (2) cases with differences larger than 2 (4) MeV. Differences are mainly seen around magic numbers 20, 28, 40 as well as in superheavy nuclei. For calculations with parameters WS1 and WS2, there are 66 cases with differences larger than 2 MeV.

It should be mentioned that in both cases studied above, the energies and deformations correspond to the minimum in the total energy surface.

\subsection{The second minimum}

There is a long history in nuclear physics studying the so-called shape coexistence and many prominent examples have been found \cite{RevModPhys.83.1467, Andreyev:2000aa, PhysRevC.90.064309}. It is also important for our study of radioactive alpha \cite{PhysRevC.90.061303} and proton decays \cite{Qi12}.
It is beyond the scope of this paper to analyze in detail all cases with possible shape coexistence. In particular, we are interested to compare the regions where shape coexistence is calculated to occur by the three different WS parameter sets. As mentioned above, in a few cases the ground state deformations predicted by the different calculations are quite different, which are related to the fact that the coexistence of low-lying states with different shapes is expected in those nuclei whereas their orders giving by different calculations are different.

\begin{figure}
\includegraphics[width=0.5\textwidth]{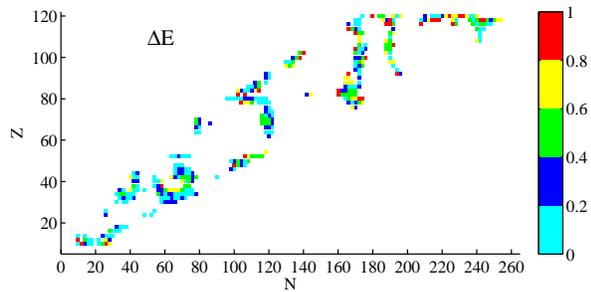}
\caption{\label{shapecuniv}(Color online)  Systematic calculations on the energy difference between the second and first deformation minima in  even-even nuclei by using the Woods-Saxon potential with WS1.}
\end{figure}

\begin{figure}
\includegraphics[width=0.5\textwidth]{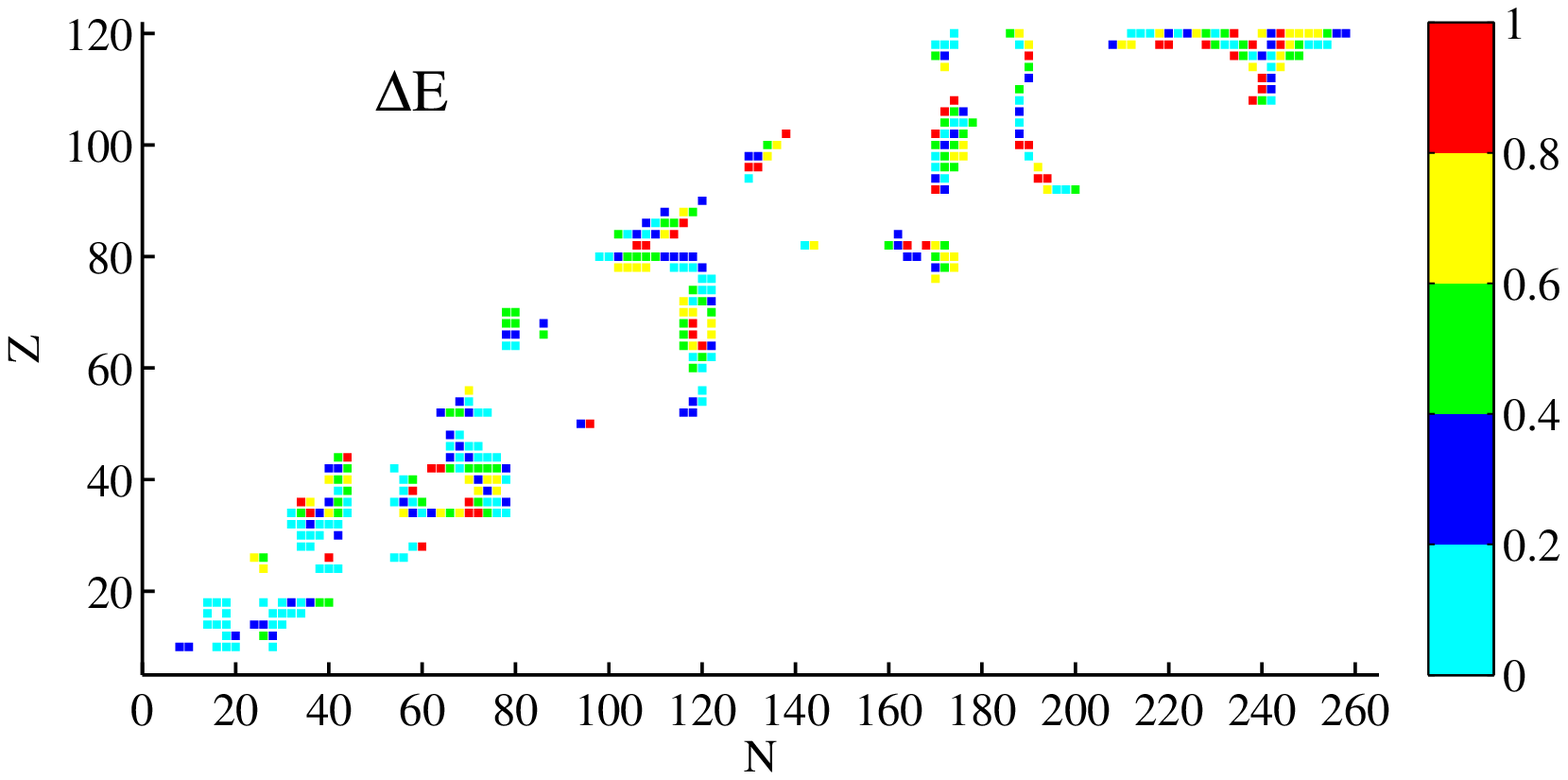}
\caption{\label{shapecrank}(Color online)  Same as Fig. \ref{shapecuniv} but for calculations with WS2.}
\end{figure}

\begin{figure}
\includegraphics[width=0.5\textwidth]{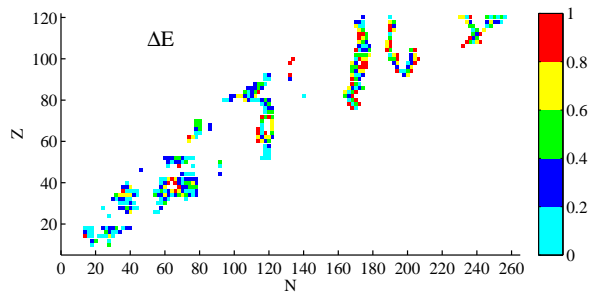}
\caption{\label{shapexu} (Color online) Same as Fig. \ref{shapecuniv} but for calculations with WS3.}
\end{figure}

In Figs. \ref{shapecuniv}, \ref{shapecrank} and \ref{shapexu} we plotted the energy difference between the second and first minima in the calculated potential energy surface for even-even nuclei over the whole nuclei chart. Only cases with energy difference smaller than 1 MeV are selected for simplicity. It can be seen from the figures that the three calculations give quite similar pattern. Shape coexistence is expected in the superheavy nuclei and nuclei around $N=120$ and 170. It may also happen in light neutron-rich nuclei as well as nuclei around $N,Z=40$.

\subsection{Effect of octupole and higher-order deformations}
Our calculations done above are restricted to the deformation space $(\beta_2, \gamma, \beta_4)$ for simplicity. Recent mass model calculations were done in the space $(\beta_2, \beta_4, \beta_6)$ in Ref. \cite{wangning101}, $(\beta_2, \gamma, \beta_4)$  in Ref. \cite{PhysRevLett.97.162502} and $(\beta_2, \beta_3, \beta_4, \beta_6)$ in Ref. \cite{MollerPRL}. 
It should be mentioned that there has a long quest studying the possible existence of static octupole and higher order correlations at nuclear ground states (see, e.g., a recent experiment \cite{nature2013} and references therein and Ref. \cite{But96} for reviews on earlier works). Both the $\beta_3$ and $\beta_6$ deformations were taken into account in the very successful mac-mic mass model calculations \cite{Moller1995185}. Multi-dimensional deformation space calculations have also been done for superheavy elements \cite{Mun01} and in recent fission studies \cite{Sta2013,mo2009}. The influence of the $\beta_6$ deformation in superheavy nuclei around $^{254}$No was also studied in Ref. \cite{Liu2014}. Those correlations may affect the results shown above.

\begin{figure}
\includegraphics[width=0.5\textwidth]{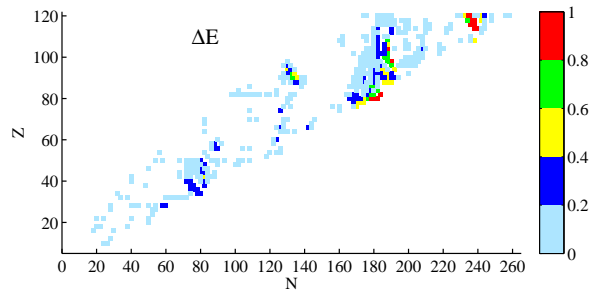}
\caption{\label{b35} (Color online) Energy differences between calculations with and without the $\beta_3$ and $\beta_5$ deformations. Those with differences smaller than 50 keV are left blank for a clearer view.}
\end{figure}

In order to understand the influence of those omitted deformation freedoms in our calculations shown above, we have done calculations within the axially symmetric deformation space $(\beta_2, \beta_3, \beta_4, \beta_5, \beta_6)$ with the WS1 parameter set. The non-axially symmetric $\gamma$ deformation is neglected due to computation limitations. In Fig. \ref{b35} we firstly compared
calculations with and without the $\beta_3$ and $\beta_5$ deformations. It is thus seen that in most cases the influence of those two deformations on the total energy is less than 200 keV. Profound $\beta_3$ deformation is seen in nuclei around $N=134$ and $Z=88$ and in superheavy nuclei around $N=180$. This is in agreement with earlier calculations \cite{Moller1995185}.

\begin{figure}
\includegraphics[width=0.5\textwidth]{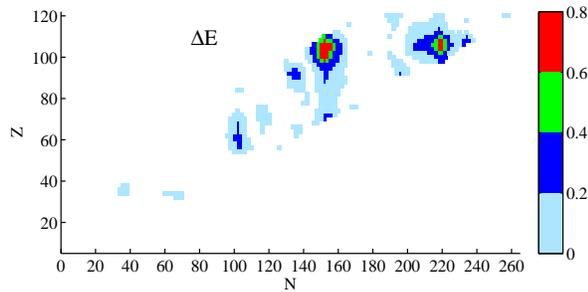}
\caption{\label{b6} (Color online) Energy differences between calculations with and without the $\beta_6$ deformation. Those with differences smaller than 50 keV are left blank.}
\end{figure}

In Fig. \ref{b6} we compared calculations with and without the $\beta_6$ deformation. The largest influence on total energy appears in nuclei around $N=152$, in agreement with Ref. \cite{Mun01}, and in superheavy elements with $N\sim220$. The large difference is around 850 keV.

\section{The total binding energy}

To explore the description power of the single-particle potentials studied above on the (negative) binding energy, we added the liquid drop energy for $E_{mac}$ \cite{Ld74,scw87}.
It is  chosen to be \cite{PhysRevC.81.044321}
\begin{eqnarray} 
E_{LDM}=a_vA+a_sA^{2/3}+ a_{sym}T\left(T+1\right)/A \nonumber\\
      + a_{syms}T\left(T+1\right)/A^{4/3}+C\frac{Z^2}{A^{1/3}}+C_4\frac{Z^2}{A} 
\end{eqnarray} 
where the terms represent the volume energy, surface energy, symmetry energy, surface symmetry energy,
Coulomb energy and correction to Coulomb energy due to surface diffuseness of 
charge distribution,  respectively. The coefficients $a_v$, $a_s$,
$a_{sym}$, $a_{syms}$, $C$ and $C_4$ are free parameters to be determined. $T=|N-Z|/2$ is the isospin. As in Refs. \cite{Xu2013247,PhysRevC.86.044323,0954-3899-42-4-045104}, the free parameters
are firstly determined by minimizing the $\sigma^2$ value in comparison with the
experimental binding energies as
\begin{eqnarray}
\sigma^2~=~\frac{1}{n}\sum_{N,Z}
      \left[ E_{calc}(N,Z) - E_{expt}(N,Z) 
\right]^2 ,
\end{eqnarray} 
where only nuclei that are heavier than $^{16}$O and have an experimental error smaller than 100 keV are considered and $n$ is the total number of data. The experimental data are taken from Ref.
\cite{1674-1137-36-12-003}. We did not consider the influence of the uncertainty induced by the errors in experimental binding energies, which is supposed to be small as compared with the unknown theoretical uncertainties \cite{0954-3899-41-7-074001}.

\begin{figure}
\includegraphics[width=0.4\textwidth]{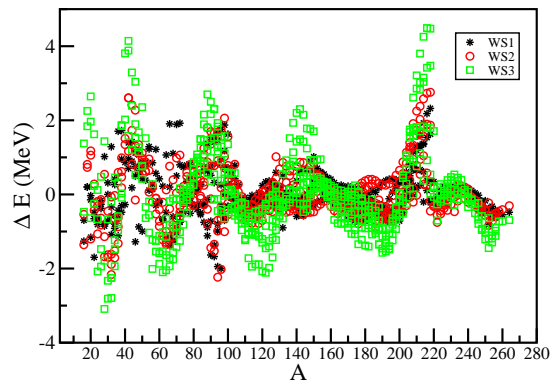}
\caption{\label{devie} (Color online) Deviations of calculations on binding energies from experimental data as a function of mass number $A$. A positive value means the binding energy is overestimated by theory.}
\end{figure}

The parameters thus fitted for the three WS calculations are given in Table \ref{tab:table1}.
For calculations with parameters WS1, one has only 2 cases with deviation between theory and experimental data larger than 2 MeV. Whereas the numbers are 14 and 44 for calculations with parameters WS2 and WS3. In all three calculations the largest deviations appear at $N=126$ isotopes $^{218}$U and $^{216}$Th. The deviations of the calculated binding energies from experimental data are plotted in Fig. \ref{devie} as a function of mass number. All three calculations, in particular those with parameters WS3, still show some kind of systematic deviations around the shell closures, which indicate that the shell correction may have not been fully taken into account.

\begin{table}
\caption{\label{tab:table1} The parameters for the liquid drop model, mean deviation and maximum deviation for the different WS calculations as determined by using the least square deviation criterion.}
\begin{ruledtabular}
\begin{tabular}{cccc}
&WS1& WS2& WS3\\
\hline
$a_v$&   -15.707  $\pm$     0.017&-15.678  $\pm$     0.017&   -15.704  $\pm$     0.017\\
 $a_s$  & 18.302  $\pm$     0.075&  18.197  $\pm$     0.075&    18.450  $\pm$     0.075\\
 $a_{sym}$ & 117.481  $\pm$     0.524& 117.431  $\pm$     0.524&   118.968  $\pm$     0.524\\
 $a_{syms}$& -161.323  $\pm$     2.990&-161.237  $\pm$     2.990&  -172.024  $\pm$     2.990\\
 $C$ &   0.717  $\pm$     0.001&   0.716  $\pm$     0.001&     0.719  $\pm$     0.001\\
 $C_4$  & -0.882  $\pm$     0.072&  -0.896  $\pm$     0.072&    -1.081  $\pm$     0.072\\
    \hline
    $\sigma$ &
    0.612&0.731 & 1.162 \\
        \hline
        Max. dev.&
2.320&    2.756 & 4.490
\end{tabular}
\end{ruledtabular}
\end{table}

\begin{figure}
\includegraphics[width=0.45\textwidth]{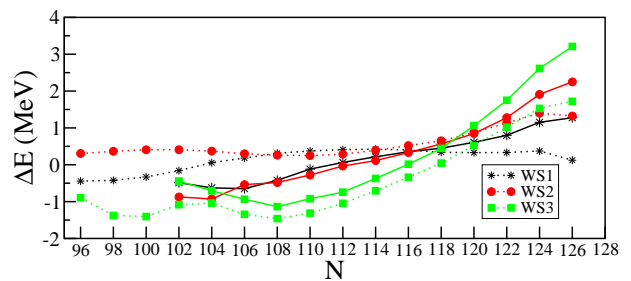}
\caption{\label{devi126}(Color online)  Same as Fig. \ref{devie} but for Pb (dotted line) and Po (solid line) isotopes below $N=126$ as a function of neutron number $N$.}
\end{figure} 
 
For binding energies calculated from WS3, in 16 cases the deviations from experimental data are larger than 3 MeV, among which one has 5 cases with $N,Z=20$ and 10 cases with $N$ around 126. For those few nuclei around $N$ or $Z=20$, the proton and neutron shell gaps predicted by the WS2 and WS3 parameter sets are much larger than those from WS1 (by about 2 MeV for neutron and 1-1.5 MeV for proton). For the nuclei with $N\sim 126$ and $Z$ between 84 and 92 with large deviations, our calculations show that the proton and neutron shell gaps predicted by the three calculations are similar to each other. However, the neutron shell correction energies predicted by the WS3 calculation are about 2 MeV more attractive than those from WS1. The results from WS2 are below those values from WS1 and WS3. Actually, as can be seen from Fig. \ref{devi126}, the deviations from experimental data show a systematic increasing trend as a function of neutron number for the different isotopic chains with neutron number below $N=126$. The increasing trend is strongest in calculations with WS3 and weakest in WS1. For all three calculations, the increasing trend is stronger in the isotopic chains above $N=82$ than that in Pb isotopes due to the different descriptions of the proton shell corrections. We notice that the single-particles energies for levels near the Fermi surface given by the WS3 and WS1 calculations are very close to each other. The difference in the shell correction energy may be due to the fact that the single-particle energies of the lowest levels are slightly deeper in WS3 than those in WS1.
 The binding energy of $^{28}$Si is underestimated by 3.09 MeV in calculations with WS3. This is related to the fact that the ground state deformations predicted by the WS2 and WS3 calculations are much larger than that from WS1, which leads to large repulsive deformation correction energies $E_{def}$ in those two calculations. On the other hand, the shell correction energies given by the three calculations are quite similar to each other.  
 
\begin{figure}
\includegraphics[width=0.4\textwidth]{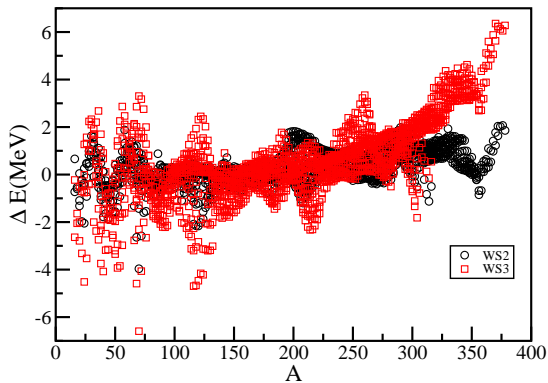}
\caption{\label{devit} (Color online) Differences between calculated binding energies with parameter set WS1 and calculations with parameters WS2 and WS3 as a function of mass number $A$. }
\end{figure}

We are particularly interested in the differences between different calculations, which may shed light on our understanding of the theoretical uncertainty. In Fig. \ref{devit} we plotted the differences between the three WS calculations as a function of $A$ for all even-even nuclei considered in this work. The largest differences between calculations with parameters WS3 and WS1 appear in the neutron-rich $^{70}$Ca and in superheavy nuclei around $Z=120$ and $N=254$. 

\begin{figure}
\includegraphics[width=0.4\textwidth]{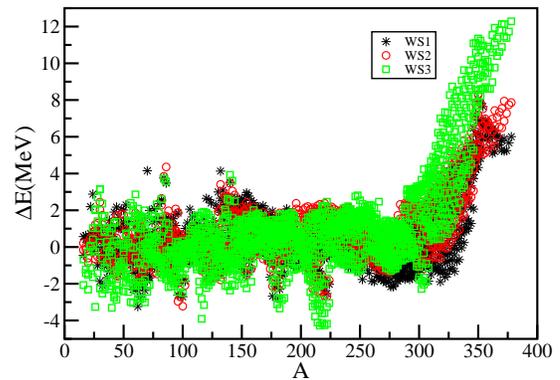}
\caption{\label{deviwang} (Color online) Differences between our mac-mic calculations on binding energies and those of Ref. \cite{wangning101} as a function of mass number $A$. }
\end{figure}

\begin{figure}
\includegraphics[width=0.4\textwidth]{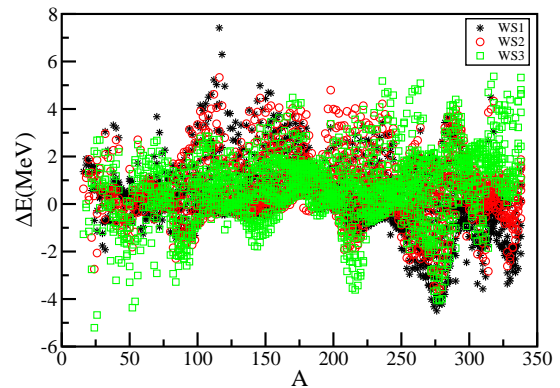}
\caption{\label{devimoller} (Color online) Differences between our mac-mic calculations on binding energies and those of Ref. \cite{Moller1995185}  as a function of mass number $A$. }
\end{figure}

In Figs. \ref{deviwang} and \ref{devimoller} we compared our calculations on the binding energies with mass formula calculations from Refs. \cite{wangning101,Moller1995185}. In the former case, systematic large deviations are seen in the superheavy region. Whereas the deviation of our calculations from those of Ref. \cite{Moller1995185} seems to be much smaller.

It may also be of interest to determine the parameters of the liquid drop model by using the so-called minimax fitting procedure.
The object of the minimax fit is to find the minimum of the maximum deviation (the least worst result) as
\begin{equation}
\varepsilon =\arg\min_{\mathbf{x}} \max_A |E_{{\rm Expt.}}(A)-E_{{\rm Calc.}}(A,\mathbf{x})|,
\end{equation}
where $\mathbf{x}$ denote the set of parameters to be determined. $\arg\min$ ($\arg\max$) stand for the argument of the minimum (maximum) for which the value of the given expression attains its minimum (maximum) value within a given set of $\mathbf{x}$. The parameters thus determined are listed in Table \ref{tab:table2}. The results given by calculations with these parameters are pretty similar to those predicted with parameters from Table \ref{tab:table1}. This is particularly the case for the WS1 parameterization. For calculations with the WS2 and WS3 parameters, the maximum deviations can be largely reduced by applying the minimax fitting criterion. However, the corresponding mean deviations increase noticeably, particularly for WS3. 

\begin{table}
\caption{\label{tab:table2} The parameters for the liquid drop model, mean deviation and maximum deviation for the different WS calculations as determined by using the minimax fitting criterion.}
\begin{ruledtabular}
\begin{tabular}{cccc}
&WS1& WS2& WS3\\ 
\hline
 $a_v$& -15.649 &   -15.672  &  -15.673\\
 $a_s$&  17.994 &    18.248  &   18.266\\
 $a_{sym}$&  115.117 &    118.397  &   116.602\\
 $a_{syms}$& -148.802 &   -172.152  &  -160.014\\
 $C$& 0.714 &   0.7161  &  0.715\\
  $C_4$&-0.769 &  -0.919  & -0.884\\
    \hline
    $\sigma$&
    0.669&0.957 & 1.650 \\
        \hline
        Max. dev. &
1.980&    2.145 & 3.365
\end{tabular}
\end{ruledtabular}
\end{table}

As can be seen from the two tables, most coefficients of the liquid drop model can be well constrained by fitting to available data except those  of the surface symmetry energy and surface correction to Coulomb energy. These two terms are quite sensitive to the choice of the different WS parameterization as well as the different fitting criteria.
This is expected as they may be more correlated with the surface properties of the nuclei induced by the nuclear shell structure.

\section{Summary} 
\label{sec:con}
We have calculated systematically the microscopic energies and nuclear deformations of even-even nuclei within the mac-mic framework with three different WS parameterizations (denoted as WS1 \cite{Dud82}, WS2 \cite{wyss} and WS3 \cite{Xu2013247}) which were constructed primarily for nuclear spectroscopy calculations. The first two WS parameterization have been previously shown to be very successful in reproducing many aspects of nuclear structure and decay properties. The simplified WS3 parameter contains an unusual SO term and was constructed mainly to explain the shell evolution in light neutron-rich nuclei  \cite{Xu2013247}. We are particularly interested to see whether the isospin dependence of the SO force has any global influence on the binding energy and on the deformation. 

It is found that the ground state deformations predicted by the three calculations are quite similar to each other. Large differences are seen mainly in neutron-rich nuclei and in superheavy nuclei. Systematic calculations on the shape-coexisting second minima are also presented. The total binding energy is estimated by adding the macroscopic energy given by the usual liquid drop model with its parameter fitted by using the least square root and minimax fitting criteria. Our calculations on the deformation and binding energy are also compared with those predicted by two available mac-mic mass formulas.

It is gratifying to notice that the WS parameters, which were fitted only to single-particle states in spherical nuclei, indeed do so well in nuclear mass calculations.
One may speculate that, on a global scale, there are smooth effects in the microscopic energy that are only depend on the bulk properties of the WS potential. 
On the other hand, it is expected that there are specific nuclear structure effects that may be sensitive to the details of the potential, e.g., the isospin dependence of the spin-orbit force and the isospin dependence of the radius.
While we describe well global properties, it is imporant to find the essentials that can differentiate a successful microscopic potential from unsuccessful ones. In other words, one can identify local properties, e.g., the emergence of new subshells and other spectroscopic properties, that can help pin down  
the sign and strength of above higher order terms of the WS potential without deteriorating its global behavior.

 It is noticed that, in a few cases in heavy nuclei, calculations with the WS3 parameter set can deviate from experimental data by as large as 4 MeV (Fig. 23). We hope that this can be improved by fine-tuning the parameterization of the potential (WS3 contains less terms than the other two parameter sets employed). We also hope that a better understanding of the theoretical uncertainties of the mac-mic model as well as the WS parameterization can be obtained.
In the future, we are also interested to see if one can pin down the isospin dependence of the SO force by fitting to binding energies as well as other properties of both stable and unstable neutron-rich nuclei.

\section*{Acknowledgement}
This work was supported by the Swedish Research Council (VR) under grant Nos. 621-2012-3805, 621-2013-4323 and the Jiangsu overseas research and training program for university prominent young and middle-aged teachers. The calculations were performed on resources provided by the Swedish National Infrastructure for Computing (SNIC) at NSC in Link\"oping and PDC at KTH, Stockholm. ZW and CQ thank  F.R. Xu for his help and N. Wang for comments.

\end{document}